\documentclass[12pt]{article}
 \usepackage[dvips]{graphicx}
\begin{document}
 \setcounter{section}{0}
\title{P-Loop Oscillator on  Clifford Manifolds and Black Hole Entropy}
\author{Carlos Castro \thanks{Center for Theoretical Studies of Physical
Systems,Clark Atlanta University,Atlanta, GA.
30314;~E-mail:castro@ts.infn.it}~~ and ~ Alex
Granik\thanks{Department of Physics, University of the Pacific,
Stockton,CA.95211;~E-mail:galois4@home.com}}
\date{}
\maketitle

\begin{abstract}
A new relativity theory, or more concretely an extended relativity
theory, actively developed by one of the authors incorporated 3
basic concepts. They are the old idea of Chew about
bootstrapping, Nottale's scale relativity, and enlargement of the
conventional time-space by inclusion of noncommutative Clifford
manifolds where all p-branes are treated on equal footing. The
latter allowed one to write a master action functional. The
resulting functional  equation is simplified and applied to the
p-loop oscillator. Its respective solution is a generalization of
the conventional point oscillator. In addition , it exhibits some
novel features:  an emergence of two explicit scales delineating
the asymptotic regimes (Planck scale region and a smooth region of
a conventional point oscillator). In the most interesting Planck
scale regime, the solution reproduces in an elementary fashion the
basic relations of string theory ( including string tension
quantization). In addition, it is shown that comparing the
massive ( super) string degeneracy with the p-loop degeneracy one
is arriving at the proportionality between the Shannon entropy of
a p-loop oscillator in D-dimensional space and the
Bekenstein-Hawking entropy of the black hole of a size comparable
with a string scale. In conclusion the Regge behavior follows
from the solution in an elementary fashion.
\end{abstract}
\section{Introduction }
Recently a new relativity was introduced \cite{cc1} -\cite{cc8}
with a purpose to develop a viable physical theory describing the
quantum "reality" without introducing by hand {\it a priori}
existing background. This theory is based upon $3$ main concepts:

1) Chew's bootstrap idea about how every $p$-brane is made of
$all$ the others and how an evolving physical system is able to
generate its own background in the process.

2) Nottale's scale relativity \cite{ln1}-\cite{ln2} which adopts
the Planck scale $\Lambda =1.62 \times 10^{-35} m$ as the minimum
attainable scale in nature.

3) a generalization of the ordinary space-time ( the concept most
important for our analysis)  by introduction of non-commutative
C-spaces leading to full covariance of a quantum mechanical loop
equation. This is achieved by extending the concepts of ordinary
space-time vectors and tensors to noncommutative Clifford
manifolds (it explains the name C-space) where all $p$-branes are
unified on the basis of Clifford multivectors. As a result, there
exists a one-to-one correspondence between single lines in
Clifford manifolds and a nested hierarchy of $0$-loop,
$1$-loop,..., $p$-loop histories in $D$ dimensions ( $D$=$p-1$)
encoded in terms of hypermatrices.

The respective master action functional $S\{\Psi [X(\Sigma)]\}$ of
quantum field theory in C-space \cite{aa1,cc4} is
\begin{equation}
\label{eq:o1}
\begin{array}{c}
S\{\Psi [X(\Sigma)]\}= \int~ [DX(\Sigma)] ~ [{1\over 2} ~({
\delta \Psi \over
 \delta X }*{ \delta \Psi \over \delta X }+ {\cal E}^2 \Psi*\Psi)+
 {g_3\over 3!} \Psi*\Psi*\Psi +\\
 \\
 {g_4\over 4!}
 \Psi*\Psi*\Psi*\Psi].
\end{array}
\end{equation}

where $\Sigma$ is an invariant  evolution parameter (a
generalization of the proper time in special relativity) such that
\begin{equation} \label{eq:o2}
\begin{array}{l}
  (d\Sigma)^2 = (d\Omega_{p+1})^2 + \Lambda^{2p}(dx_\mu dx^\mu)+
  \Lambda^{2(p-1)}d\sigma_{\mu\nu}d\sigma^{\mu\nu}+...\\
  +(d\sigma_{\mu_1\mu_2...\mu_{p+1}}d\sigma^{\mu_1\mu_2...\mu_{p+1}})
  \end{array}
 \end{equation}

\begin{equation}
\label{eq:o3}
 X(\Sigma)=\Omega_{p+1}I+\Lambda^p x_\mu \gamma^\mu+
 \Lambda^{p-1}\sigma_{\mu\nu}\gamma^\mu\gamma^\nu+...
\end{equation}

is a Clifford algebra-valued line "living" on the Clifford
manifold outside space-time, $\Lambda$ is the Planck scale that
allows to combine objects of different dimensionality in Eqs.(2,3)
and the multivector $X$ Eq.(\ref{eq:o3}) incorporates both a
point history given by the ordinary ( vector) coordinates $x_\mu$
and the holographic projections of the nested family of all
$p$-loop histories onto the embedding coordinate spacetime
hyperplanes : $\sigma_{\mu\nu},....\sigma_{\mu_1
\mu_2...\mu_{p+1}}$. The scalar ( from the point of view of
ordinary Lorentz transformations but $not$ from the {\bf C}-space
point of view ) $\Omega_{p+1}$ is the invariant proper
$p+1=D$-volume associated with a motion of a maximum dimension
$p$-loop across the $p+1 =D$-dim target spacetime. Since a
Cliffordian multivector with $D$ basis elements ( say, $e_1,
e_2,...,e_D$) has $2^D$ components, our vector $X$ has also $2^D$
components.

Generally speaking, action (\ref{eq:o1}) generates a master
Cantorian (strongly fractal) field theory with a braided Hopf
quantum Clifford algebra. This action is unique in a sense that
the above algebra selects terms allowed by the action. The
quadratic terms  are the usual kinetic and mass squared
contributions; the cubic terms are the vertex interactions; the
quartic terms are the braided scattering of four Clifford lines.
In what follows we restrict our attention to a truncated version
of the theory by applying it to a linear $p$-loop oscillator.

This truncation is characterized by the following $3$
simplifications. First, we dropped nonlinear terms in the action,
that is the cubic term ( corresponding to vertices) and the
quartic (braided scattering) term. Secondly, we freeze all the
holographic modes and keep only the zero modes which would yield
conventional differential equations instead of functional ones.
Thirdly, we assume that the metric in C-space is flat.

\section{Linear Non-Relativistic p-loop Oscillator}

We begin this section by properly defining what one means by
"relativistic" from the point of view of the new relativity. The
complete theory is the master field theory whose action functional
admits a noncommutative braided quantum Clifford algebra. As a
result of the postulated simplifications, we are performing a
reduction of the field theory to an ordinary quantum mechanical
theory. It must be kept in mind that fields are $not$ quantized
wave functions. For this reason the wave equations that we will be
working with refer to a nonrelativistic theory in {\bf C}-spaces.

Hence, using all these restrictive assumptions, we obtain from the
action (\ref{eq:o1}) a C-space p-loop wave equation for a linear
oscillator
\begin{equation}
\label{eq:o4}
\begin{array}{l}
  \{-{1\over2}{1\over\Lambda^{p-1}}[\frac{\partial^2}{\partial
 {x_\mu}^2}+\Lambda^2\frac{\partial^2}{(\partial\sigma_{\mu\nu})^2}+
 \Lambda^4\frac{\partial^2}{(\partial \sigma_{\mu\nu\rho})^2}+...+
 \Lambda^{2p}\frac{\partial^2}{(\partial\Omega_{p+1})^2}]+\\
 \\
 {m_{p+1}\over2}{1\over L^2}[\Lambda^{2p}x{_\mu}^2+\Lambda^{2p-2}
{\sigma_{\mu\nu}}^2+...+\Omega^2_{p+1}]\}\Psi = T\Psi
 \end{array}
\end{equation}

where $\frac{\partial^2}{(\partial x_\mu)^2} = g^{\mu\nu}
\frac{\partial}{\partial x^\mu}\frac{\partial}{\partial x^\nu},
\hspace{2mm}  \frac{\partial^2}{(\partial
\sigma_{\mu\nu})^2}=G^{\mu\nu\rho\tau}\frac{\partial} {\partial
\sigma^{\mu\nu}}\frac{\partial}{\partial\sigma^{\rho\tau}}, ...,
etc,$   $G^{\mu\nu\rho\tau}$ is some suitably symmetrized product
of the two ordinary metric-tensors $g^{\mu\nu} g^{\rho\tau}$, $T$
is tension of the spacetime-filling $p$-brane, $ D = p+1$,
$m_{p+1}$ is the parameter of dimension $(mass)^{p+1}$ ,
parameter $L$ (to be defined later) has dimension $length^{p+1}$
and we use units $\hbar = 1, c = 1$. A generalized correspondence
principle {\footnote{In the limit of $\Lambda/a \rightarrow 0$
volume $ \Omega_{p+1}$, holographic coordinates $\sigma_{\mu\nu},
\sigma_{\mu\nu\rho}, ...\rightarrow 0$, and $p$-loop oscillator
should become a point oscillator, that is $p$-loop histories
collapse to a point history}} allows us to introduce the
following qualitative correspondence between the parameters
$m_{p+1} , L$, and mass $m$ and amplitude $a$ of a point
(particle) oscillator:

\begin{center}
        $m_{p+1}("mass") \longleftrightarrow m$,

\smallskip
 $   L("{amplitude}") \longleftrightarrow a$
 \end{center}

We rewrite Eq.(\ref{eq:o4}) in  the dimensionless form as follows
\begin{equation}
\label{eq:o5}
 \{ \frac{\partial^2}{\partial \tilde{x}_{\mu}^2}+ \frac{\partial^2}{\partial
\tilde{\sigma}_{\mu\nu}^2}+...-
(\tilde{\Omega}^2+\tilde{x}_{\mu}^2+\tilde{\sigma}_{\mu\nu}^2+...)+
 2{\cal{T}}\}\Psi =0
\end{equation}
where ${\cal{T}}=(T/\sqrt{{\cal{A}}}m_{p+1})$ is the dimensionless
tension and ${\cal A}$ is a scaling parameter that will be
determined below.
\begin{center}
$   \tilde {x}_{\mu}\hspace{1mm} =
  {\cal{A}}^{1/4}\frac{\Lambda^p}{L}x_\mu,\hspace{3mm}
    \tilde{\sigma}_{\mu\nu}\hspace{1mm} = {\cal{A}}^{1/4}
    \sigma_{\mu\nu}\frac{\Lambda^{p-1}}{L},...,\tilde{\Omega}_{p+1}={\cal{A}}^{1/4}\frac{\Omega_{p+1}}{L}$
\end{center}
are the dimensionless arguments, $\tilde{x}_{\mu}$ has ${C_D}^1
\equiv D$ components, $\tilde{\sigma}_{\mu\nu}$ has ${C_D}^2
\equiv \frac{D!}{(D-2)!2!}$ components, etc.

Inserting the usual Gaussian solution for the ground state into
the wave equation (4) we get the value of ${\cal A}$ :
\begin{center}
${\cal{A}} \equiv {m_{p+1}L^2}/{\Lambda^{p+1}}$
\end{center}
Without any loss of generality we can set ${\cal{A}}=1$ by
absorbing it into $L$. This will give the following geometric
mean relation between the parameters $L, \hspace{1.5mm}m_{p+1}$,
and $\Lambda$

$$L^2 = {\Lambda^{p+1}}/{m_{p+1}} \Rightarrow \Lambda^{p+1} < L
<   { 1 \over m_{p+1} }   $$ meaning that there are three scaling
regimes. The scale represented by generalized Compton wavelength
$  (1/ m_{p+1})^{( 1/p+1) }    $ will signal a transition from a
smooth continuum to a $fractal$ ( but continuous ) geometry. The
scale $L$ will signal both a  $discrete$ and $ fractal$ world (
like El Naschie's Cantorian-fractal spacetime models and $p$-adic
quantum mechanics ) and $\Lambda$ the quantum gravitational
regime.

The dimensionless coordinates then become

\bigskip

$\tilde{x}_\mu =
\sqrt{\Lambda^{p+1}m_{p+1}}\hspace{1mm}x_\mu/{\Lambda},~~
\tilde{\sigma}_{\mu\nu}
=\sqrt{\Lambda^{p+1}m_{p+1}}\hspace{1mm}\sigma_{\mu\nu}/\Lambda^2,...,$\\

$\tilde{\Omega}_{p+1}=\sqrt{\Lambda^{p+1}m_{p+1}}\hspace{1mm}\Omega_{p+1}/
\Lambda^{p+1}$

\bigskip
The dimensionless combination $\Lambda^{p+1}m_{p+1}$ indicating
existence of two separate scales : $\Lambda$ and
$(1/{m_{p+1}})^{\frac{1}{p+1}}$ obeys the following double
inequality:
\begin{equation}
\label{eq:o6} \sqrt{{m_{p+1}\Lambda^{p+1}}} < 1 <
\sqrt{\frac{1}{m_{p+1}\Lambda^{p+1}}}
 \end{equation}
Relations (\ref{eq:o6}) define two asymptotic regions:

1)the "discrete-fractal" region characterized by
${m_{p+1}\Lambda^{p+1}} \sim 1$, or the Planck scale regime, and

\smallskip
2)the "fractal/smooth phase transition ", or the low energy region
characterized by ${m_{p+1}\Lambda^{p+1}}<< 1$.

\bigskip
Since the wave equation (\ref{eq:o5}) is diagonal in its arguments
( it is separable ) we

represent its solution  as a product of separate functions of
each of the dimensionless arguments
$\tilde{x}_{\mu},\tilde{\sigma}_{\mu\nu},..., $
\begin{equation}
\label{eq:o7}
  \Psi =\prod _{i}F_i(\tilde {x}_i)\prod_{j<k}F_{jk}(\tilde\sigma_{jk})...
\end{equation}

Inserting (\ref{eq:o7}) into (\ref{eq:o5}) we get for each of
these functions the Whittaker equation:
\begin{equation}
\label{eq:o8}
    Z^{\prime\prime} - (2{\cal{T}} - \tilde {y}^{2})Z =0
    \end{equation}
where $Z$ is any function $F_i, F_{ij},...$, $\tilde{y}$ is the
respective dimensionless variable $\tilde{x}_{\mu}$,$\tilde
{\sigma}_{\mu\nu},...$, and there  are all in all $2^D$ such
equations. The bounded solution of (\ref{eq:o8}) is expressed in
terms of the  Hermite polynomials $H_n(\tilde{y})$
\begin{equation}
\label{eq:o9}
    Z \sim e^{-\tilde{y}^2/2}H_n(\tilde{y})
        \end{equation}
Therefore the solution to Eq.({\ref{eq:o5}) is
\begin{equation} \label{eq:o10}
 \Psi \sim
 exp[-({\tilde{x}_\mu}^2+{\tilde{\sigma}_{\mu\nu}}^2+...+\tilde{\Omega}^2_{p+1})]
 \prod_{i}H_{n_i}(\tilde{x}_i)\prod
 _{jk}H_{n_{jk} }(\tilde{\sigma}_{jk})...
 \end{equation}
where there are $D$ terms corresponding to $n_{1},n_{2},...,
n_{D}$, $D(D-1)/2$ terms corresponding to $n_{01}, n_{02},...,$
etc. Thus the total number of terms corresponding to the $N$-th
excited state $(N
=n_{x1}+n_{x2}+...+n_{\sigma_{01}}+n_{\sigma_{02}}+...)$  is
given by the degree of the Clifford algebra in $D$ dimensions
$2^D$.

The respective value of the tension of the $N$-th excited state is
\begin{equation}
\label{eq:o11}
 T_N= (N + \frac{1}{2} 2^D) m_{p+1}
\end{equation}
yielding quantization of tension.

Expression (\ref{eq:o11}) is the analog of the respective value
of the $N$-th energy state for a point oscillator. The analogy
however is not complete. We point out one substantial difference.
Since according to a new relativity principle \cite{cc1}
-\cite{cc8} all the dimensions are treated on equal footing
(there are no preferred dimensions) all the modes of the $p$-loop
oscillator( center of mass $x_\mu$, holographic modes, $p+1$
volume) are to be excited collectively. This behavior is in full
compliance with the principle of polydimensional invariance by
Pezzaglia \cite{wp1}. As a result, the first excited state is not
$N=1$ ( as could be naively expected) but rather $N=2^D$.
Therefore
\begin{center}
    $T_1 \rightarrow T_{2^D} = \frac{3}{2}(2^D m_{p+1}$)
\end{center}
instead of the familiar $(3/2)m$.

Recalling that $L$ is analogous to the amplitude $a$ and using
the analogy between energy $E \sim m\omega^2 a^2$ and tension $T$,
we get $T = m_{p+1}\Omega^2L^2$. Inserting this expression into
Eq.({\ref{eq:o11}) we arrive at the definition of  the "frequency"
$\Omega$ of the $p$-loop oscillator:
\begin{equation}
\label{eq:o12}
    {\Omega_N} =\sqrt{(N +  2^{D-1})
    \frac{m_{p+1}}{\Lambda^{p+1}}}
\end{equation}

where we use $L =\sqrt{\Lambda^{p+1}/m_{p+1}}$.

Having obtained the solution to Eq.(\ref{eq:o5}), we consider in
more detail the two limiting cases corresponding to the above
defined 1) fractal and 2) smooth regions. The latter ( according
to the correspondence principle) should be described by the
expressions for a point oscillator. In particular, this means that
the analog of the zero slope limit in string theory, the field
theory limit, is a collapse of the $p$-loop histories  to a point
history :

$$ \Lambda \rightarrow 0,~~ m_{p+1} \rightarrow \infty,~~ T
\rightarrow \infty,~~ \sigma_{\mu\nu},
\sigma_{\mu\nu\rho},......        \rightarrow   0,~~~ L
\rightarrow 0.~ ...$$ and these limits are taken in such a way
that the following combination reproduces the standard results of
a point-particle oscillator :

\begin{equation}\label{eq:o13}
\tilde{x}_{\mu}=
\frac{x_{\mu}}{\Lambda}\sqrt{{m_{p+1}\Lambda^{p+1}}} \rightarrow
x_{\mu}/a
\end{equation}
where the $nonzero$  parameter $a > \Lambda$ is a finite
quantity  and is nothing but the amplitude of the usual
point-particle oscillator !  In string theory, there are two
scales, the Planck scale $\Lambda$ and the string scale $ l_s >
\Lambda$. Without loss of generality we can assign $ a \sim l_s$.
A large value of $a >> \Lambda$ would correspond to a
``macroscopic''  string. We shall return to this point when we
address the black-hole entropy.

Using Eq.(\ref{eq:o13}) we find ${m_{p+1}}$ in terms of the other
variables :
 \begin{equation}
 \label{eq:o14}
 {m_{p+1}}\rightarrow{(M_{Planck})}^{p+1}
 (\frac{\Lambda}{a})^2 < {(M_{Planck})}^{p+1}
 \end{equation}
where the Planck mass $M_{Planck} \equiv 1/\Lambda$. Notice that
in the field-theory limit, $\Lambda \rightarrow 0$, when the loop
histories collapse to a point-history, Eq.(14) yields $  {m_{p+1}}
\rightarrow \infty$ as could be expected. From Eqs.({\ref{eq:o11})
and (\ref{eq:o12}) follows that in this region
\begin{equation}
\begin{array}{l}
\label{eq:o15}
 \ {T_N} \sim
 (M_{Planck})^{p+1}(\frac{\Lambda}{a})^2< (M_{Planck})^{p+1}\\
\ {\Omega_N} \sim
(\omega_{Planck})^{p+1}\frac{\Lambda}{a}< (\omega_{Planck})^{p+1}\\
\omega_{Planck}=1/\Lambda
\end{array}
\end{equation}
in full agreement with  this region's scales as compared to the
Planck scales.

At the other end of the spectrum ( discrete-fractal/quantum
gravity  region) where
 $m_{p+1}\Lambda^{p+1}\sim 1$ we would witness a collapse of all
the scales to only one scale, namely the Planck scale $\Lambda$.
In particular, this means that the string scale $a \sim l_s \sim
\Lambda$, and the oscillator parameters become
\begin{equation}
\label{eq:o16}
 \ \tilde{x}_{\mu}= \frac{x_{\mu}}{\Lambda}
\sqrt{\Lambda^{p+1}m_{p+1}} \sim
\frac{x_\mu}{\Lambda},\hspace{2mm} m_{p+1} \sim
\frac{1}{\Lambda^{p+1}} \equiv (M_{Planck})^{p+1},
\end{equation}

The ground state tension is :

$$T_o \sim m_{p+1} \sim\frac{1}{\Lambda^{p+1}}$$ These relations
are the familiar relations of string theory. In particular, if we
set $p=1$ we get the basic string relation
\begin{center}
$ T \sim \frac{1}{\Lambda^2} \equiv \frac{1}{\alpha'}$
\end{center}

\smallskip
Above we got two asymptotic expression for $m_{p+1}$
\[ m_{p+1} = \left\{\begin{array}{ll}
\ \Lambda^{-(p+1)}(\Lambda/a)^2 & \mbox{if~
$\Lambda/a < 1$}\\
\ \Lambda^{-(p+1)} & \mbox{if ~$m_{p+1}\Lambda^{p+1}\sim 1,~a
\sim \Lambda$}
  \end{array}
  \right. \]
It is suggestive to write $m_{p+1}\Lambda^{p+1}$ as power series
in $(\Lambda/a)^2$ (e.g., cf. analogous procedure in
hydrodynamics \cite{DV1}):
\begin{center}
$m_{p+1}\Lambda^{p+1}\equiv F(\Lambda/a)
=(\frac{\Lambda}{a})^2[1+ \alpha_1
(\frac{\Lambda}{a})^2+\alpha_2(\frac{\Lambda}{a})^4+...]$
\end{center}
where the $small $ coefficients  $\alpha_i$ are such that the
series is convergent for $a \sim \Lambda$.

For example, the dimensionless coordinate ${\tilde x}_\mu$ given
by Eq.(13), becomes in the field theory limit $\Lambda \rightarrow
0$, after performing a Taylor/binomial expansion of the square
root :

$${\tilde x}_\mu = {x^\mu \over a} [ 1 + {\alpha_1 \over 2}
({\Lambda \over a})^2 +.....] \rightarrow  {x^\mu \over a}$$
Notice that $a$ is a finite nonzero quantity.

\smallskip
If $p=1 (p+1 = D = 2) $ then  for the ground state $N =0$ Eq.(11)
yields the ground energy per unit string length :
$T_{ground}=2m^2$ (see footnote\footnote{that is for a point
oscillator we get
$E_{ground}=\hbar\omega/2=\sqrt{T_{ground}/8}$}). Returning to the
units $\hbar$, and introducing $1/a = \omega$ ( where $\omega$ is
the characteristic frequency) we get (cf.ref \cite{cc5})
\begin{center}
$\hbar_{eff} = \hbar \sqrt{1+ \alpha_1
(\frac{\Lambda}{a})^2+\alpha_2(\frac{\Lambda}{a})^4+...}$
\end{center}
Truncating the series at the second term , we recover the string
uncertainty relation [5] :
\begin{center}
$ \Delta x \Delta p >  { 1\over 2 }| [ x , p ]|  = {\hbar_{eff}
\over 2} \Rightarrow  \Delta x >  {\hbar \over 2 \Delta p} +
\beta  {\Lambda^2 \over \hbar } \Delta p $ \end{center} where
$\beta$ is a multiplicative parameter. As $\Lambda \rightarrow 0$
one recovers the ordinary Heisenberg uncertainty relations.
Interestingly enough, the string uncertainty relation until
recently did not have " a proper theoretical framework for the
extra term" \cite{ew1}. On the other hand, this relation has
emerged as one of the results of our theoretical model [5] .

As a next step we find the degeneracy associated with the $N$-th
excited level of the $p$-loop oscillator. The degeneracy $dg(N)$
is equal to the number of partitions of the number $N$ into a set
of \hspace{1mm}  $2^D$ numbers $$N = \{n_{x^1} + n_{x^2}+ ...+
n_{x^D}+ n_{\sigma_{\mu\nu}}+ n_{\sigma_{\mu\nu\rho}}+ ...+ n_{
\Omega_{ p+1} } \}.$$ This means that there is a $collective$
center of mass excitations, holographic area,
volume,...excitations  given by the quantum numbers $
n_{x^D};n_{\sigma_{\mu\nu}},....... n_{ \Omega_{ p+1} } $
respectively. These collective $extended$  excitations are the
$true$ quanta of a background independent quantum gravity. Thus
the degeneracy is

\begin{equation}
\label{eq:o17}
 dg(N)= \frac{\Gamma(2^D +N)} {\Gamma(N+1)\Gamma({2^D})}
\end{equation}
where $\Gamma$ is the gamma function.

We compare $dg(N)$  (\ref{eq:o17}) with the asymptotic quantum
degeneracy of a massive (super) string state given by Li and
Yoneya \cite{l1}:
\begin{equation}
\label{eq:o18}
  dg(n)= exp \hspace{2mm}[2\pi \sqrt {n\frac{d_s-2}{6}}\hspace{2mm}]
\end{equation}
where $d_s$ is the string dimension and $n >> 1$. To this end we
equate (\ref{eq:o18}) and degeneracy (\ref{eq:o17}) of the first
excited state ( $N=2^D$) of the $p$-loop. This could be justified
on physical grounds as follows. One can consider different frames
in a new relativity: one frame where an observer sees strings only
(with a given degeneracy) and another frame where the same
observer sees a collective excitations of points, strings,
membranes,p-loops, etc. The results pertinent to the degeneracy
(represented by a number) should be invariant in any frame.

Solving the resulting equation $ dg(N)=  dg(n)$ with respect to
$\sqrt{n}$ we get
\begin{equation}
\label{eq:o19}
 \sqrt{n} = \frac{1}{2\pi}{\sqrt\frac{6}{d_s-2}}\hspace{2mm}{Ln[
 \frac{\Gamma(2^{D+1})}{\Gamma(2^D+1)\Gamma(2^D)}]}
 \end{equation}
 The condition $n >> 1$ implies that $D >> 1$ thus simplifying
(\ref{eq:o19}). If we set $d_s = 26$ ( a bosonic string) and use
the asymptotic representation of the logarithm of the gamma
function for large values of its argument
\begin{center}
$Ln\Gamma(z) =Ln(\sqrt{2\pi})+(z-1/2)Ln(z)- z + O(1/z)$
\end{center}
we obtain the following logarithm of the degeneracy yielding the
entropy :
\begin{equation}
\label{eq:o20}
 Entropy = Ln [ dg(n)] \sim \sqrt{n} \approx 2^D \hspace{1mm}\frac{ln(2)}{2\pi} \sim 2^{D-1}\sim N
 \end{equation}
From (Eq.\ref{eq:o18}) follows that for $n >>1$ the $entropy =
Ln~[dg(n)] \sim \sqrt{n}$. Let us consider a Schwarzschild black
hole whose Schwarzschild radius $R$
\begin{center}
$R \sim (GM)^{1\over (d-3)}$
\end{center}
The black hole mass $M$ and the string length $l_s \sim R$ obey
the Regge relation $$ l^2_s M^2 \sim R^2M^2 = n $$ which implies
that the world sheet area and mass are quantized in Planck units :
$l^2_s = {\sqrt n} \Lambda^2$ and $ M^2 = {\sqrt n} M^2_{Planck} =
{\sqrt n}\Lambda^{-2}$.  Li and Yoneya \cite{l1} obtained the
following expression for the Bekenstein-Hawking entropy of a
Schwarzchild black hole of a radius  $R \sim l_s$ \footnote {It
should be mentioned that the linear relation between the black
hole entropy and is justified only for a narrow region of
dimensions $ D \sim [4,5]$; in general, this relation loses its
linear character \cite{cc8,cc9}} : $$ S_{BH} \sim { A \over G }
\sim { R^{d-2}\over G} \sim { (GM)^{d-2 \over d-3} \over G} = G^{
{1\over d-3}  } M^{{d-2 \over d-3}} =RM. $$

From the last two equations Li and Yoneya deduced that  the
$(d-2)$- dimensional horizon area in Planck units was  $S_{BH}$
is $$S_{BH} \sim \sqrt{n}.$$

Now taking into account Eq. (\ref{eq:o20}) we obtain
\begin{equation}
\label{eq:o21}
 S_{BH} \sim 2^{D-1}
 \end{equation}

This is a rather remarkable fact: the Shannon entropy of a
$p$-loop oscillator in $D$-dimensional space ( for a sufficiently
large $D$), that is a number $N=2^D$  ( the number of bits
representing all the holographic coordinates), is proportional to
the Bekenstein-Hawking entropy of a Schwarzschild black hole. For
a more rigorous study of the connection between Shannon's
information entropy and the quantum-statistical (thermodynamical)
entropy see the work of Fujikawa \cite{fuj}. Because light is
trapped inside, the Black Hole horizon is also an information
horizon.

To summarize , expression(\ref{eq:o20}) allows us to easily
compare it with the Regge behavior of a string spectrum for large
values of $n
>>1$. To this end we associate with each bit of a $p$-loop
oscillator a fundamental Planck  length $\Lambda$, with area
$\Lambda^2$ , mass $ 1/\Lambda$, etc. The macroscopic string
length is of the same order as the Schwarzschild radius, $R^2
\sim l_s^2 \sim Area_s ~= ~$N$ \times \Lambda^2,~ {m_s}^2~ = ~$N$
\times M_{Planck}^2$. On the other hand, according to
(\ref{eq:o20}) $N$ $\sim \sqrt{n}$ which yields
\begin{center}
$l^2_s \sim \sqrt{n}\hspace{1mm}\Lambda^2; \hspace{3mm} m_s^2 \sim
\sqrt{n}\hspace{1.5mm} M_{Planck}^2$
\end{center}
Therefore using the Regge relation between angular momentum and
mass-squared, the respective angular momentum $J$ in units where
$\hbar = 1$  is of the order :
\begin{center}
$ J = m^2 \times\hspace{1mm} {l_s}^2 \sim n M_{Planck}^2\Lambda^2
= n$
\end{center}
where we use $M_{Planck} \Lambda ~ \equiv ~ 1$ by definition. One
can derive the Regge relation more precisely using the
Bohr-Sommerfield  correspondence principle in quantum mechanics
but applied to the string case in question:

$$ Action = \int~ P_{\mu\nu} d\Sigma^{\mu\nu}\sim (Tension) (
Area) \sim n \hbar. $$ For more details of this relationship we
refer to [1,3,5] where we have shown that the area-momentum
variable  $ P_{\mu\nu}$, conjugate to the area tensor
$\Sigma^{\mu\nu}$,  obeys the Hamilton-Jacobi equation similar to
the point particle case.

Addressing the black holes we still encounter the main remaining
question: where does Einstein gravity appear in all of this? The
answer lies in the behavior of a self-gravitating gas of loops.
This is precisely where the Bohr correspondence limit operates.
The large $ n >>1$ limit is similar to the Bohr correspondence for
the hydrogen atom ( highly excited energy states merge with the
continuum ) where the product of $ n \hbar$ remains finite when $
n \rightarrow \infty; \hbar \rightarrow 0$.  As the size of the
string gets larger, the $p$-loop oscillator begins to resemble a
gas of strings, or more precisely  a gas of string-bits, a
string-polymer. As it gets even larger, the correspondence limit
comes into play, and the gas of loops will begin to gravitate. The
derivation of Einstein equations for this self-gravitating gas of
loops in the large $n$ limit : the Einstein tensor equals stress
energy tensor ( with/without a cosmological constant ) will be
the topic of a future publication.

\section{Conclusion}
Application of  a simplified linearized equation derived from the
master action functional of a new ( extended) relativity to a
$p$-loop oscillator has allowed us to elementary obtain rather
interesting results. First of all, the solution explicitly
indicates existence of $2$ extreme regions characterized by the
values of the dimensionless combination $m_{p+1}\Lambda^{p+1}$ :

\smallskip
 $1)$ the fractal region where  $m_{p+1}\Lambda^{p+1}\sim
1$ and $2$ scales collapse to one, namely Planck scale $\Lambda$

and

$2)$ the smooth region where $m_{p+1}\Lambda^{p+1}<< 1$ and we we
recover the description of the conventional point oscillator.
Here $2$ scales are present , a characteristic "length" $a$  that
we identified with the string scale $l_s$ and the ubiquitous
Planck scale $\Lambda$ ( $a > \Lambda$) thus demonstrating
explicitly the implied validity of the quantum mechanical
solution in the region where $a/\Lambda > 1$.

For a specific case of $p=1$ (a string) the solution yields ( once
again in an elementary fashion) one of the basic relations of
string theory $T = 1/\alpha'$. In addition, it gives us  a string
uncertainty relation ( this time derived), which in turn is a
truncated version of a more general uncertainty relation obtained
earlier \cite{cc5}.

Comparing the degeneracy of the $first~collective$ state of the
p-loop for a very large number of  dimensions $D$ with the
respective expressions for the massive ( super) string theory
given by Li and Yoneya \cite{l1} we found that the Shannon
entropy ( which is also in agreement with the logarithm of the
degeneracy of states ) of a $p$-loop oscillator in
$D$-dimensional space ( for a sufficiently large $D$), that is a
number $N=2^D$  ( the number of bits representing all the
holographic coordinates), is proportional to the
Bekenstein-Hawking entropy of the Schwarzschild black hole.

The Regge behavior of the string spectrum for large $n>>1$ also
follows from the obtained solution thus indicating its, at least
qualitatively correct, character. Thus a study of a simplified
model ( or "toy") problem of a linearized $p$-loop oscillator
gave us ( with the help of elementary calculations) a wealth of
both the well-known relations of string theory ( usually obtained
with the help of a much more complicated mathematical
technique)and some additional relations ( the generalized
uncertainty relation). This indicates that the approach advocated
by a new relativity might  be very fruitful, especially if it
will be possible to obtain analytic results on the basis of the
full master action functional leading to functional nonlinear
equations whose study will involve braided Hopf groups.
\bigskip

{\bf Acknowledgements } The authors would like to thank
E.Spallucci and S.Ansoldi for many valuable discussions and
comments.


\begin{thebibliography}{99}
 \bibitem{cc1} C. Castro , " Hints of a New Relativity Principle from
    $p$-Brane Quantum Mechanics "  J. Chaos, Solitons
    and Fractals {\bf 11}(11)(2000) 1721
\bibitem{cc2} C. Castro ,    " The Search for the Origins of $M$ Theory : Loop
    Quantum Mechanics and Bulk/Boundary Duality " hep-th/9809102
 \bibitem{cc3} S. Ansoldi, C. Castro, E. Spallucci , " String Representation of
Quantum Loops " Class. Quant. Gravity {\bf 16 }   (1999)
1833;hep-th/9809182
\bibitem{cc4}C.Castro, " Is Quantum Spacetime Infinite Dimensional?"  J. Chaos, Solitons and
Fractals{\bf 11}(11)(2000) 1663
\bibitem{cc5} C. Castro,  " The String Uncertainty Relations follow
from the  New Relativity Theory " hep-th/0001023; Foundations of
Physics , to be published
\bibitem{cc6} C. Castro, A.Granik,  "On $M$ Theory, Quantum Paradoxes
    and the New Relativity " physics/ 0002019;
\bibitem{cc7}C. Castro, A.Granik,  "How a New Scale Relativity Resolves
   Some Quantum Paradoxes", J.Chaos, Solitons, and Fractals {\bf
   11}(11) (2000) 2167.
 \bibitem{cc8} C. Castro, "An Elementary Derivation of the Black-Hole
    Area-Entropy Relation in Any Dimension " hep-th/0004018
\bibitem{ln1} L. Nottale, "Fractal Spacetime and Microphysics :Towards a
    Theory of Scale Relativity " World Scientific, 1993;
\bibitem{ln2} L. Nottale, "La Relativite dans tous ses Etats " Hachette Literature, Paris, 1998.
\bibitem{aa1} A. Aurilia, S. Ansoldi , E. Spallucci,   J. Chaos,
 Solitons and Fractals. {\bf 10}(2-3)  (1999) 197 .
 \bibitem{wp1} W.Pezzaglia, "Dimensionally Democrtic Calculus and
 Principles of Polydimensional Physics", gr-qc/9912025
 \bibitem{DV1} M.Van Dyke, "Perturbations Methods in Fluid
 Mechanics", Academic Press, NY, London (1964)
 \bibitem{ew1} E.Witten, "Reflections on the Fate of
 Spacetime",Physics Today (April 1996) 24
\bibitem{cc9} C.Castro, A.Granik, "Scale Relativity in Cantorian
$\cal E^\infty$ Space and Average Dimensions of the World",
J.Chaos,Solitons and Fractals (2000)
\bibitem{l1} M.Li and T.Yoneya, "Short Distance Space-Time
Structure and Black Holes",J. Chaos, Solitons and Fractals. {\bf
10}(2-3)  (1999) 429
\bibitem{fuj} K.Fujikawa, "Shannon's Statistical Entropy and the H-theorem in
Quantum Statistical Mechanics", cond-mat/0005496


\end{thebibliography}
\end{document}